\documentclass[preprint,aps,draft]{revtex4}
\usepackage{graphicx}
\usepackage{dcolumn}
\usepackage{bm}
\usepackage{amssymb}
\usepackage{amsmath}
\usepackage{amsfonts}
\usepackage[]{graphicx}

\begin{document}

\title{The End of the Many-Worlds?\\
  (or Could we save Everett's interpretation)}

\author{Aur\'{e}lien~Drezet}
\email{aurelien.drezet@neel.cnrs.fr} \affiliation{Institut N\'eel UPR 2940, CNRS-University Joseph Fourier, 25 rue des Martyrs, \\
38000 Grenoble, France}

\date{\today}

\begin{abstract}

\end{abstract}

 \maketitle
\section{Introduction}
\indent The aim of this text is to discuss in a rather critical way
the so called many worlds interpretation (MWI) proposed originally
by H. Everett in 1957~\cite{Everett}. I am a proponent of the pilot
wave interpretation (PWI) defined by L. de Broglie in 1927 and D.
Bohm in 1952~\cite{Holland} and while MWI shares with PWI a strong
commitment for determinism there are also fundamental differences
between the two approaches. The most important one concerns probably
how both theories attempt to interpret quantum probability within
their own ontological frameworks and this will be the topic of the
present chapter. As I will show however MWI is more difficult to
accept than  PWI in the sense that it has dramatic problems with its
ontology which cannot be ignored if we want to interpret
probabilities.\\
\indent First, let us say a few words about ontology. MWI and PWI
are realistic approaches to quantum mechanics. This means that they
are trying to introduce a clean ontological structure into the
formalism of quantum mechanics.  What is indeed so extraordinary
about quantum mechanics is that the formalism appeared in 1925-1927
without a clear ontological framework in complete opposition with
classical approaches.  Instead, N. Bohr, W. Heisenberg, M. Born and
others managed to develop a pragmatic and instrumentalist
interpretation in which only macroscopic apparatuses and detectors
possess a `clear' definition. This way of thinking, very much in
harmony with the dominant positivism of this time, relies on the
need to consider as existing only what is seen by an `observer'
(i.e. the so called  quantum observable). In classical physics this
positivistic approach was already introduced by the philosopher E.
Mach, with his phenomenalistic philosophy of science, and by the
chemist W. Oswald with his strong criticism of atomism (i.e., in his
debate with L. Boltzmann). The logical positivism of M. Schlick, R.
Carnap, P. Franck, and H. Reichenbach perpetuated this methodology
in the XX's century and considered that ontology is pure metaphysics
and should be removed of any positive science (actually they are
strongly mistaken: any science is necessarily metaphysical on its
theoretical ground as shown already by D. Hume. Theory is a pure
creation of the human mind and needs to be tested with experimental
facts. This is the basis of the hypothetico-deductive method which
was in particular defended by L. Boltzmann and later by A.~Einstein
in his debated with W.~Heisenberg and N.~Bohr). However, Bohr's way
of thinking is not completely identical to these various positivist
approaches. For Bohr the problem is mainly experimental and is
associated with the  existence of the quantum of action $\hbar$.
Indeed, the classical ontologies based on waves and particles are
not able to give a clear unambiguous picture of quantum reality as
shown for instance by the famous wave-particle duality paradox.
Therefore, one should try a different minimalist approach in which
these classical concepts, while necessary for the empirical
description of phenomena, do not have the same ontological values as
in the classical world. The Copenhagen interpretation says something
like that: Take the Schrodinger equation
\begin{equation}i\hbar\frac{\partial}{\partial
t}|\Psi\rangle_t=\hat{H}|\Psi\rangle_t
\end{equation} with its wave function $|\Psi\rangle_t$ and its hamiltonian
$\hat{H}$. Don't try to `see' a physical propagation of a wavy
thing; use it as a practical tool for defining a probability
obtained by a `classical observer'. The observer, sentient or not,
(detector or automaton are part of the interpretation as well)
possesses a well defined position in space and time and is therefore
foreign to the quantum formalism. Quantitative statements are
introduced into the theory through Born's probability rule which
reads
\begin{equation} P_\alpha(t)=|\langle\alpha|\Psi_t\rangle|^2
\end{equation} where $P_\alpha$, the probability of the outcome
$\alpha$ (associated with the observable operator $\hat{A}$) , is
related to the quantum state by squaring the norm of the amplitude
$c_\alpha(t)=\langle\alpha|\Psi_t\rangle$ (where we have the
eigenvalue relation $\hat{A}|\alpha\rangle=\alpha|\alpha\rangle$).
Ontological questions about what happens to the system between the
preparation and the measurement are beyond the scope of Bohr's
interpretation. This answer is fine for an experimentalist in his
lab and can be used with confidence for all practical purposes
(FAPP). However, this methodology keeps open some unsolved
fundamental questions. Questions such as  the meaning of the wave
function for the Universe, or the Schrodinger cat, or  Wigner's
Friends, or the Heisenberg shifty split, and so on and so forth are
very pertinent and they cannot simply be escaped by labeling them
metaphysical. After all, the quantum wave theory was developed by
Schrodinger before Bohr gave his interpretation and there is no
reason why the Copenhagen method should be the only pertinent one.
Actually, the situation is even a bit ironical since Bohr's
interpretation came only of late.  De Broglie proposed his own
`double solution' theory already in 1925, before the Schrodinger
equation was discovered (de Broglie was using the Klein-Gordon
equation which he invented for his own purpose). Schrodinger
proposed his own theory in 1926. In fact de Broglie's view contains
as a sub-product the PWI while Schrodinger's, if correctly
interpreted in the configuration space has as a necessary
consequence MWI (as is recognized by Everett himself in his PhD
thesis~\cite{Everett}).
\section{Everett and Bohm}
\indent Now, I will try to define briefly the ontology of MWI and
PWI. I will start with PWI. In PWI the wave function
$|\Psi\rangle_t$ has an ontological meaning independently of the
existence or nonexistence of the observer. $|\Psi\rangle_t$ is an
actual guiding wave in a real quantum field for particles which are
somehow surfing on the wave. Here I will be rather conservative and
consider only the non-relativistic case for a single particle of
mass $m$ without magnetic potential. The velocity of the point like
particle in PWI is given by
\begin{equation}
\mathbf{v}(t)=\frac{d}{dt}\mathbf{x}(t)=\frac{\hbar}{m}\textrm{Im}[\frac{\boldsymbol{\nabla}
\psi(\mathbf{x}(t),t)}{\psi(\mathbf{x}(t),t)}].
\end{equation}
This guidance formula is enough for solving paradoxes such as
wave-particle duality by defining a complex trajectory for the
particle. The wave carries the particle into regions where the field
is non vanishing and omits regions where the quantum field cancels,
therefore offering a contrasting view to the Bohr-Heisenberg dictum
that such kind of representation should be prohibited (in the same
way the famous `no-go' theorem by von Neumann~\cite{von} vanishes
into pure smoke~\cite{Holland,Bellvon}). Alternatively, we can
introduce a quantum potential $Q_\Psi$ acting on the particle and
modifying the newtonian force induced by the external potentials
$V(\mathbf{x},t)$~\cite{Holland}. In the configuration space for
many particles $\mathbf{x}_1(t)$,$\mathbf{x}_2(t)$ etc... the theory
is highly nonlocal and can be used to solve the Einstein Podolsky
Rosen paradox (EPR)\cite{EPR} in full agreement with Bell's
theorem~\cite{Bell,Holland}. I point out that I was deliberately
quite unprecise and vague concerning the nature of the quantum field
$\Psi$. Indeed, in PWI,   $\Psi$ guides the particles but there is
no reaction of the particle on the wave. It is for this reason
better to wait for a better understanding of the ontology of the
wave function in the future. Probably, PWI, if it survives, will
have to be modified or completed by a more satisfying theory in
which particles and fields (manifested through waves) will be
dynamically connected. This was the hope of both de Broglie and Bohm
with different strategies and we should not here comment further.
Anyway, even if such a theory would exist one day, this doesn't mean
that the current PWI will not be correct anymore in the same way
that Newton's theory of gravitation is not completely invalidated by
general relativity. Additionally, as it was pointed out many times
by de Broglie there is a strong analogy between the status of $\Psi$
in the PWI and the action $S(q,t)$ in the old Hamilton-Jacobi
theory. Therefore, we will here only consider $\Psi$ as an effective
field or a
`nomological entity' keeping its understanding for future works. \\
\indent Now, of course PWI makes sense only if we can introduce
probabilistic elements into the theory in order to explain Born's
formula Eq.~2.  Schrodinger's equation contains enough material to
do that unambiguously. Indeed, from the local current conservation
law $\boldsymbol{\nabla}\cdot\mathbf{J} +\partial_t \rho=0$ with
$\mathbf{J}(\mathbf{x},t)=\mathbf{v}(t)\rho(\mathbf{x},t)$ and
\begin{equation}
\rho(\mathbf{x},t)=|\psi(\mathbf{x},t)|^2,
\end{equation}
it is obviously `natural' to interpret $\rho(\mathbf{x},t)$ as a
density of presence for the particle located at $\mathbf{x}$ at time
$t$.  The local conservation plays the role of Liouville theorem's
in statistical mechanics. Therefore, if $\rho(\mathbf{x},t)$ is
effectively interpreted as a density of probability of presence
$P(\mathbf{x},t)$ at a given time the interpretation will still be
valid at any other times in agreement with the conservation law.
This was essentially the reasoning of de Broglie in 1927 and of Bohm
in 1952. However, one can try to go further and attempt to justify
the probability law $P(\mathbf{x},t)=|\psi(\mathbf{x},t)|^2$ with
other assumptions. This will be very similar to modern statistical
mechanics trying to justify the  microcanonical, or canonical
Boltzmann-Gibbs ensemble with deeper reasonings. Several attempts
have been done in the recent years focussing on either H-theorem,
coarse graining, Boltzmann-Typicality, and deterministic chaos (for
a review see \cite{Callender}).  I will not discuss that further
since my aim was only to point that PWI is very similar to classical
physics, no better no worse. The difficulties and questionings about
equilibrium and non-equilibrium cited by Boltzmann are now
translated in the PWI framework into the question of how to justify
the uniqueness of quantum equilibrium and how to reach such an
equilibrium. The solutions are probably very similar for the
classical and quantum cases (but the role of entanglement and non
locality is still not so clear) so that solving the problem in
classical statistical
mechanics would give a strong insight on the quantum PWI version.\\
\indent Let us now turn to  MWI. As said before, MWI is mainly the
strict application of Schrodinger's original theory in the
configuration space. In MWI, $\psi$ is an ontological field entity
as in  PWI. But here there is not particle at all only a continuous
wave! For Schrodinger in the case of the single electron, the  wave
function $e|\psi(\mathbf{x},t)|^2$ describes a local electron charge
density not a probability.  The quantum field $\psi(\mathbf{x},t)$
in such a theory is somehow very similar to the classical
electromagnetic field in the old fashioned Maxwell's theory, or to
the metric tensor in general relativity. In recent years,
practitioners and proponents of MWI, in particular L. Vaidman, often
called $|\psi(\mathbf{x},t)|^2$ a measure of existence or a degree
of reality~\cite{Vaidman} (the definition of L.~Vaidman is more
general of course but it agrees with mine if by existing quantum
`world' we mean a state like $|\mathbf{x}\rangle$. However since
there is no preferred basis in the Hilbert space the choice is
arbitrary). This is clearly reminiscent of Schrodinger's language
about charge density. Of course, if we think of the electron as a
delocalized charge distribution, we get the obvious objections
already given by Bohr and Heisenberg that a wave packet is spreading
through space and time and that this could not explain the
experimental facts where electrons are appearing as localized grains
or quanta on detectors. In a diffraction or interference experiment
the electron would go through both holes and could not create a
singular event on the screen. This is the well known historical
reason why the Schrodinger solution was quickly abandoned already in
1927. Furthermore, observe that already for a single electron wave
function the spatial coordinate representation doesn't have any
particular status in the theory.  We could also have chosen a
momentum representation $|\mathbf{k}\rangle$ instead of
$|\mathbf{x}\rangle$ this would not have modified the information
contained into the quantum state. Therefore, there is no privileged
basis in the MWI of Schrodinger and Everett in the same way that it
is not more or less physical to consider the Fourier transform
$\tilde{E}(k)$ of an electromagnetic field as more or less real than
the field $E(x)$ itself. However, we have something new in quantum
physics. A superposition such as $|\textrm{here}\rangle+
|\textrm{there}\rangle$ for a single electron means that two wave
packets localized in the 3D space are superposed. Since however
experiments show that an electron is only
detected here OR there, it seems to us that MWI should have fundamental difficulties in dealing with this reality.\\
\indent However, Everett resurrected Schrodinger's theory by
introducing entanglement in the many particles configuration space.
His hope was that entanglement added to the Schrodinger idea would
solve completely the measurement paradox and all other
contradictions. We will see that this is not so easy. To start with
entanglement, consider an ideal two-particle Universe. In the
coordinate representation and at a given time $t$, the wave function
for the system is written $\psi(\mathbf{x}_1,\mathbf{x}_2,t)$. What
is here the meaning of $\mathbf{x}_1$, and $\mathbf{x}_2$? For a
single electron $\mathbf{x}$ is a spatial coordinate labeling the
point in the 3D space.  But here we have two points of coordinates
$\mathbf{x}_1$, and $\mathbf{x}_2$. This means that if we want to
interpret $\psi(\mathbf{x}_1,\mathbf{x}_2,t)$ as an ontological
field we should extend a bit the classical framework of field
theory. Generally, in classical field theory we consider only local
fields defined at a single position. The field itself can obey a
nonlinear equation in order to create for example solitons or
`bunched' field. Here however, the quantum field $\psi$ require two
coordinates for its definition $\mathbf{x}_1$, and $\mathbf{x}_2$.
This involves a new form of non-locality or wholeness while the
equation (i.e Eq.~1) is kept strictly linear. How to call such a
field? Maybe `web-field' could be a good alternative name to wave
function since it actually represents a correlation between two or
many points in space and time. The theory can of course be extended
to the relativistic domain  using the Tomonaga-Schwinger equation
for the functional $\Psi[\sigma]$ where $\sigma$ is a space-like
hyper-surface in the 4D Universe. Now, how  will this solve the
paradox of the spreading electron which was a dead-end for the old
Schrodinger interpretation? Consider again our wave function
$\psi(\mathbf{x}_1,\mathbf{x}_2,t)$. Suppose at the initial time the
system factorizes, i.e.,
$\psi(\mathbf{x}_1,\mathbf{x}_2,t)=\psi_1(\mathbf{x}_1,t)\phi_2(\mathbf{x}_2,t)$.
We can take  $\psi_1(\mathbf{x}_1,t)$ as a freely expanding wave
packet diffracted by an external potential like a double slit screen
(the fact that the screen or more generally an external potential is
a classical concept is here irrelevant and it is there only to
simplify the discussion). The second wave function
$\phi_2(\mathbf{x}_2,t)$ is associated with a localized wave-packet
confined in, for example, a Coulomb potential, let's says, in a
energy level $E_0$.   When the first electron collides with the
second the system can exchange energy and momentum and this
represents some information transfer to be used in a basic
measurement protocol. In other worlds, the modification of the
second electron state will lead to entanglement and non-locality.
Consider the final state
$\sum_i\psi_i(\mathbf{x}_1,t)\phi_i(\mathbf{x}_2,t)$ where the sum
is over possible outcomes. If we consider only a two state system
for $\phi_i$ then we have here a basic electron detector plate with
the ground state meaning undetected while the excited states are
recorded. Of course, we could also introduce a third electron
localized in state $\chi_3(\mathbf{x}_3,t)$ before the interaction
and actually factorized from the rest of the wave function. After
the interaction we will obtain a state like
\begin{equation}
\psi'_1(\mathbf{x}_1,t)\phi_g(\mathbf{x}_2,t)\chi_{e}(\mathbf{x}_3,t)+\psi''_1(\mathbf{x}_1,t)\phi_e(\mathbf{x}_2,t)\chi_{g}(\mathbf{x}_3,t)+...
\end{equation} where the dots indicate other terms irrelevant for the present purpose (for example the states where electrons 2 and 3 are not disturbed by the interaction).
Clearly, we have here an entanglement representing a basic spatial
correlator (the equivalent in optics would be an Hanbury Brown and
Twiss apparatus). If electron 2 is in the excited state then
electron 3 is in the ground state and reciprocally, this is `10' or
`01' information associated with two different bits. In the language
of Bohr's interpretation this would mean that an observer can only
detect a particle once since there is only one electron 1. However,
here we are not in the Copenhagen instrumentalist framework but in
the MWI. Our Universe contains only three electrons and no observer
at all. Still, with Everett's web-function Eq.~5 represents an
ontological field and the difference between
$|e_2\rangle|g_3\rangle$ and $|g_2\rangle|e_3\rangle$ is completely
real in an objective sense (for example those states are orthogonal
in the Hilbert space). However, now comes the fundamental issue:
since there is no collapse how should we interpret the superposition
given by Eq.~5 in  MWI? A state like $\psi_1'|g_2\rangle|e_3\rangle+
\psi_1''|e_2\rangle|g_3\rangle$ could also equivalently be written
$\frac{1}{2}(\psi_1'+\psi_1'')(|g_2\rangle|e_3\rangle+|e_2\rangle|g_3\rangle)+\frac{1}{2}(\psi_1'-\psi_1'')(|g_2\rangle|e_3\rangle-|e_2\rangle|g_3\rangle)$.
Since there is no observer able to collapse the state this
equivalent representation shows that we don't have yet our
macroscopic world where electrons appear either here or there but
not at both places. While entanglement allowed us to describe a
measuring device in quantum mechanics (i.e., it constitutes an
example of the shifty split of the Heisenberg cut) it didn't
apparently remove our problem in the MWI. In other words, the fact
that we considered more and more electrons didn't solve the problem
it only propagated it to a larger system thanks to entanglement.
Still, it was Everett's hope that entanglement could somehow solve
the measurement issue: how could that be ? Everett's strategy was to
consider larger and larger systems until we can consider a conscious
being or at least machines or robots sufficiently sophisticated to
have memory sequences of recorded events. The hope was that at a
certain scale to be defined the `collapse' (i.e. in the language of
the MWI the `split') should occur. This is what we should now
analyze. Considering the previous example, i.e., Eq.~5, we see that
entanglement will indeed propagate to any other systems in
interaction with our detectors. If the detector is emitting a signal
going to a larger system able to memorize, i.e., to modify in a
rather stable way a chemical, or molecular arrangement, in a
mechanical `brain' we can imagine entangled state such as
\begin{equation}
\psi'_1(\mathbf{x}_1,t)|\ddot{\smile}\rangle+\psi''_1(\mathbf{x}_1,t)|\ddot{\frown}\rangle+...
\end{equation}
with obvious notations for the states of the brain. I point out that
the `ket' notation for the brain is a compact way of speaking about
$\langle X_1,X_2,...,X_N|\ddot{\smile}\rangle$ where $N\sim10^{23}$
is the number of particles in the brain (including its environment,
the detectors, and ultimately all the Universe). However the
representation chosen is here irrelevant. Furthermore, we have here
obviously $\langle\ddot{\frown}|\ddot{\smile}\rangle=0$ since the
unitarity of the quantum evolution should preserve the orthogonality
of the pointer states present in  Eq.~5 (i.e. $\langle
g_2,e_3\rangle| e_2,g_3\rangle=0$). The brain states are of course
physical states since with Everett and von Neumann we accept
`psychophysical parallelism' \cite{Everett,von} that is the
functionalist view whereby the mind supervenes on the brain like
software on hardware. Everett's strong belief was that the state of
awareness or consciousness of the observer or robot is a new
essential ingredient in the theory. However, don't forget that
Everett believed strongly in functionalism and that for him the
introduction of minds in quantum mechanics was very different from
what was later assumed by believers in the so called many-minds
interpretation(s)~\cite{Albert} in which `minds' different of brain
states (i.e. without obvious supervenience relations with the
hardware) played a fundamental role as well. I will not discuss such
alternative approaches here since the abandonment of the
psychophysical
parallelism is definitely too much for me.\\
\indent Going back to Eq~6, we see that these two awareness states
exist and evolve as if they were alone. The reason is that they
constitute two independent solutions of Eq.~1 (I will develop that
important point below). But as we said before unitarity allows many
representations of Eq.~6 like for example
\begin{eqnarray}
\frac{(\psi'_1+\psi''_1)}{\sqrt{2}}\frac{(|\ddot{\smile}\rangle+|\ddot{\frown}\rangle)}{\sqrt{2}}
+\frac{(\psi'_1-\psi''_1)}{\sqrt{2}}\frac{(|\ddot{\smile}\rangle-|\ddot{\frown}\rangle)}{\sqrt{2}}+...
\end{eqnarray}
Here we have superposed `cat' states
$|\ddot{\smile}\rangle\pm|\ddot{\frown}\rangle$ the meaning of which
is unclear. What is the `feeling' of  a brain in such a cat state ?
Is this not a fatal problem for MWI? Are we not introducing
furtively a new axiom favoring a representation, i.e., a preferred
basis at the
detriment of not preserving full unitarity?\\
\indent However we point out that for Everett the awareness basis is
not so much privileged but better considered as `special' or
`particular' in the sense that we are always free to use a different
basis if we wish. Therefore, unitarity is not violated. Still, to be
convincing, the MWI should explain why
$|\ddot{\smile}\rangle\pm|\ddot{\frown}\rangle$ is not an awareness
state. May be there are actually such mental states like
$|\ddot{\smile}\rangle\pm|\ddot{\frown}\rangle$ but that we can not
feel how it is to be like them. Maybe the question is a bit like
asking how it feels like to be a bee or a cat. Maybe not. This is a
bit magic or mysterious for many and it is probably why some tried
to shift from  the MWI  to  the many-minds interpretation(s). Of
course, if the two brain states evolve independently there is no
apriori reason to mix them but what is more is that we can
apparently use an argumentation based on inter-subjectivity and
entanglement to see why we never meet people in such cat states.
Indeed, if I met You in the street, going out of the lab, we could
create an entangled state like
\begin{equation}
\psi'_1(\mathbf{x}_1,t)|\ddot{\smile}_{\textrm{Me}},\ddot{\smile}_{\textrm{You}}\rangle+\psi''_1(\mathbf{x}_1,t)|\ddot{\frown}_{\textrm{Me}},\ddot{\frown}_{\textrm{You}}\rangle+...
\end{equation}
Therefore, there has to be an inter-subjective agreement between
awareness states of Me and You. Moreover, like for Eq.~6 and 7,
there are also states like
$|\ddot{\smile}_{\textrm{Me}},\ddot{\smile}_{\textrm{You}}\rangle\pm|\ddot{\frown}_{\textrm{Me}},\ddot{\frown}_{\textrm{You}}\rangle$
but now they involve Me, You and all the part of the Universe having
interacted with us. There is thus an inter-`subjective' agreement
between the cat states since Eq.~8 can be written like
\begin{eqnarray}
\frac{(\psi'_1(\mathbf{x}_1,t)+\psi''_1(\mathbf{x}_1,t))}{\sqrt{2}}
\frac{(|\ddot{\smile}_{\textrm{Me}},\ddot{\smile}_{\textrm{You}}\rangle+|\ddot{\frown}_{\textrm{Me}},\ddot{\frown}_{\textrm{You}}\rangle)}{\sqrt{2}}\nonumber\\
+\frac{(\psi'_1(\mathbf{x}_1,t)-\psi''_1(\mathbf{x}_1,t))}{\sqrt{2}}
\frac{(|\ddot{\smile}_{\textrm{Me}},\ddot{\smile}_{\textrm{You}}\rangle-|\ddot{\frown}_{\textrm{Me}},\ddot{\frown}_{\textrm{You}}\rangle)}{\sqrt{2}}...
\end{eqnarray}
but you should not bother about this since neither Me nor You are
feeling such cat states separately.\\
\indent However, this argumentation is not completely convincing
 to everyone (see for example Penrose~\cite{penrose}) since there are
apparently other counterintuitive ways to write Eq.~8. For
instance what about writing Eq.~8 like
\begin{eqnarray}
\frac{(\psi'_1(\mathbf{x}_1,t)|\ddot{\smile}_{\textrm{Me}}\rangle+\psi''_1(\mathbf{x}_1,t)|\ddot{\frown}_{\textrm{Me}}\rangle)}{\sqrt{2}}
\frac{(|\ddot{\smile}_{\textrm{You}}\rangle+|\ddot{\frown}_{\textrm{You}}\rangle)}{\sqrt{2}}\nonumber\\+\frac{(\psi'_1(\mathbf{x}_1,t)|\ddot{\smile}_{\textrm{Me}}\rangle-\psi''_1(\mathbf{x}_1,t)|\ddot{\frown}_{\textrm{Me}}\rangle)}{\sqrt{2}}
\frac{(|\ddot{\smile}_{\textrm{You}}\rangle-|\ddot{\frown}_{\textrm{You}}\rangle)}{\sqrt{2}}+...
?
\end{eqnarray} The cat state for You are now involved while I am mixed with the first electron state.
The well known problem with such an expression is that it is not
obvious and univocal to factorize the Me and You like I did (this is
the reason why I wrote Me and You inside the same ket vector in
Eqs.~8, 9). Indeed, the two protagonists are strongly interacting
with a complex environment.  The possibility to separate  Me from
You is therefore in large part arbitrary and in practice physically
impossible in the lab. Where to put the border between Me and You is
a bit a question of taste. This is clearly reminiscent of the
Wigner's friend paradox: If before interacting with my friend I am
in the state
$|\ddot{\smile}_{\textrm{Me}}\rangle+|\ddot{\frown}_{\textrm{Me}}\rangle$
it is hoped  that after meeting we should be in a state  like
$|\ddot{\smile}_{\textrm{Me}},\ddot{\smile}_{\textrm{You}}\rangle+|\ddot{\frown}_{\textrm{Me}},\ddot{\frown}_{\textrm{You}}\rangle$
and that possible state factorization doesn't make any sense. Also,
for the same reason  a state like
$|\ddot{\smile}_{\textrm{Me}},\ddot{\sim}_{\textrm{You}}\rangle+|\ddot{\frown}_{\textrm{Me}},\ddot{\sim}_{\textrm{You}}\rangle$
can not be easily factorized as
$(|\ddot{\smile}_{\textrm{Me}}\rangle+|\ddot{\frown}_{\textrm{Me}}\rangle)|\ddot{\sim}_{\textrm{You}}\rangle$
since the border separating Me from You is fuzzy and shifty. We
would like to obtain some kind of superselection rules~\cite{Joos}
here in order to prove that such strange cat states are forbidden in
nature. For such reasons, many proponents in the MWI like
D.~Zeh~\cite{Zeh} or D.~Wallace and S.~Saunders~\cite{Everett,book},
following also W. Zurek~\cite{book}, often emphasize that
decoherence should be
taken into account in this argumentation. \\
\indent Decoherence is the averaging of an operator over the
`irrelevant' degrees of freedom associated with the
environment~\cite{Zeh}.  The idea is that a macroscopic quantum
object like a brain should be considered as an open system
interacting with its environment. By averaging over degrees of
freedom associated with this environment we can transform a pure
system into a mixture characterized by for example rate equations.
Overlap terms between different possible environments will decay in
time very quickly so that we can use a mixture instead of a pure
evolution in the Hilbert space. This is a nice way for removing
quantum interferences from a reduced evolution and it is reminiscent
of the famous which-path-experiment in the Young double slit setup.
Decoherence supposes already that we can interpret $|\psi|^2$ as a
probability in agreement with Born's rule Eq.~2. This creates
problems for both Bohr's and Everett's interpretations but not for
the PWI. Indeed, for Bohr the observer or the apparatus is a key
ingredient. But, tracing over some degrees of freedom means that the
environment is actually needed as an observer in this theory (`the
environment is watching you'). This is an amendment to the original
Copenhagen interpretation but this is not actually so dramatic since
the theory is instrumentalist so that improving on it does not
really touch the problem of ontology. This problem doesn't exist at
all for the PWI since probabilities have a classical meaning here.
Tracing over the environment will only introduce a supplementary
emerging ignorance like in thermodynamics. For the MWI however,
there is no yet probability so that decoherence cannot be
interpreted in the same way. We have only degrees of reality and
measure of existence so that tracing actually means summing over
some of these degrees of reality. How can this help the MWI? It
seems rather to create additional problems. The answer is not so
clear because in my opinion proponents of the MWI didn't yet reach a
consensus. One point which is often emphasized is that the
privileged basis should be very robust in the sense that only in
this basis  is the system immune to the interaction with the
environment. This means in particular that if an experimentalist was
able to prepare and isolate during even a short period a
superposition like $|\ddot{\smile}\rangle\pm|\ddot{\frown}\rangle$
the system should decohere very fastly, i.e., in a time probably
smaller than $10^{-20} s$~\cite{Tegmark}. While this result is
generally interpreted using probability it has also an absolute
meaning as a measure-theoretical way  to define effective
orthogonality between
independent observer branches.\\
\indent In this context the ideas of
D.~Wallace~\cite{Everett,book,pattern} on `patterns' are very
interesting for the MWI. What is a pattern? Referring to
D.~Dennett's work, Wallace interprets a pattern like a cat or a
tiger as an emerging structure inside the quantum evolution. The
border of an emerging structure is not always clearly defined at the
macroscopical scale since the number of atoms is huge. Including
some atoms in You or Me is a bit arbitrary so that we generally
don't  care about this shifty border between observers. We already
discussed this problem before and we should develop this a bit more.
Rigorously, we should call a pattern any complete solution of Eq.~1
for a given Hamiltonian. Referring again to the single electron
problem if $|\textrm{here}(t)\rangle$ and
$|\textrm{there}(t)\rangle$ are separate solutions and therefore
patterns then the sum $|\textrm{here}(t)\rangle+
|\textrm{there}(t)\rangle$ is also representing a viable pattern of
the theory. However, the linearity doesn't destroy the individual
patterns `here' and `there' and in that sense it really means that
we have two localized electrons here. In the same sense
$|\ddot{\smile}\rangle$ and $|\ddot{\frown}\rangle$ are representing
patterns if they are independent solutions of the full evolution
(including the environment). The linear superposition Eq.~6 is also
an allowed pattern so that we have really two `Me' here:  one happy
and one unhappy (Wallace calls that favoring multiplicity over
superposition~\cite{pattern}). Moreover, while the orthogonality of
the states was here a consequence of the orthogonality of the
pointer states this is not necessary. Two solutions of Eq.~1 can be
viable patterns without being orthogonal.  Still, in practice
decoherence will ensure that
$\langle\ddot{\frown}|\ddot{\smile}\rangle\approx 0$ is a good
approximation after a very short time due to the complexity of the
environment~\cite{Joos} (neglecting Poincar\'{e}'s recurrence over
the finite time of  human or Universe existence) and we could say
that decoherence somehow create and select emerging patterns
behaving classically (i.e., without interference) from the full
spectrum of possible solutions (those solutions that have
no good thermodynamical properties are not considered).\\
\indent There is an important point to emphasize here: considering a
single photon of energy $E$ interacting with a beam splitter, we
generally say that we end up with two independent possibilities
$|\textrm{reflected}\rangle$ and $|\textrm{transmitted}\rangle$
which are added and represent two independent pattern solutions of
the free hamiltonian.  This is however an approximation since the
boundary conditions at the beam splitter make these two `solutions'
inseparable. Only the sum $|\textrm{reflected}\rangle+
|\textrm{transmitted}\rangle$ is a solution of the time independent
scattering problem. Still, for practical purposes we are dealing
with  finite wave packets well localized in space and time. This
means that after interacting with the external potential (the beam
splitter) we can approximately and asymptotically consider the two
solutions $|\textrm{reflected}\rangle$ and
$|\textrm{transmitted}\rangle$ as independent and evolving freely.
So when we say that we end up with two independent branches this is
already an approximation even with very simple systems and without
including  macroscopic decoherence. I mention that because it is
also trivially true for Eqs.~6, and 8 and patterns like
$\psi'_1(\mathbf{x}_1,t)|\ddot{\smile}\rangle+\psi''_1(\mathbf{x}_1,t)|\ddot{\frown}\rangle$
or
$\psi'_1(\mathbf{x}_1,t)|\ddot{\smile}_{\textrm{Me}},\ddot{\smile}_{\textrm{You}}\rangle+\psi''_1(\mathbf{x}_1,t)|\ddot{\frown}_{\textrm{Me}},\ddot{\frown}_{\textrm{You}}\rangle$
are not actually  made of two independent sub-solutions but only
constitute complete and inseparable solutions of the full Universal
evolution. If you think about that  then  you realize that indeed
$|\ddot{\smile}\rangle$ and $|\ddot{\frown}\rangle$ cannot in
general be rigorously independent patterns albeit seen as emerging
and approximate structures whose condition
$\langle\ddot{\frown}|\ddot{\smile}\rangle\approx 0$ is very robust,
 due to decoherence. Of course, cat patterns, as given in Eqs.~7
and 9, are not exact solutions as well and we cannot really separate
them from the full evolution. This at least shows that the old
debates about a preferred basis in the MWI is a bit empty. However,
if patterns have only an approximative meaning, they also let the
question about the meaning of cat states
$|\ddot{\smile}\rangle\pm|\ddot{\frown}\rangle$ unanswered. It is
probably better for these reasons to return to our provocative
answer: To be a in a state like
$|\ddot{\smile}\rangle\pm|\ddot{\frown}\rangle$ is perhaps  a bit
like feeling as a bee or a cat. Maybe it is here with us  during all
our life. Maybe not. The ontology of the MWI is definitely very
strange and debates about its self-consistency will certainly
continue during many years in many-worlds.
\section{probability in the Many-Worlds
interpretation} \indent Well, if this is enough for the ontology in
the MWI interpretation, what about probability? This is the weaker
(worst?) point in the theory and it stirred up so much emotional
debates within the years that it could be too long to summarize all
points and argumentations here. Additionally, since decoherence
needs a definition of probability to work, it seems that introducing
probability could also help the ontological problem in the MWI. Now,
Everett introduced probability in his MWI in two ways. The first way
is actually predating Gleason's theorem~\cite{Gleason} which Everett
discovered independently in a simplified version. I will not
summarize this well known reasoning but just remind that its aim is
to find the most plausible measure $\mu(\Psi)$ in the Hilbert space
which could represent a probability of occurence. The result, using
some natural assumptions about linearity, is that the most natural
measure is the one given by Born's law, i.e., Eq.~2. Still, this is
not yet a probability but just the proof that if we are going to
attribute a probability to the $\Psi$ state then Born's law is the
best choice~\cite{Everett}. In the recent years
D.~Deutsch~\cite{Deutsch}, D. Wallace and S.
Saunders~\cite{Everett,book} on the one side and  W.~Zurek on the
other side~\cite{Zurek,book} tried to find an alternative
mathematical demonstration using some other symmetries (called
envariance by Zurek, and decision theory by Deutsch) than those
considered by Gleason and Everett. The reasoning is that due to
entanglement we can in a natural and mathematical way give a precise
statement for Laplace's principle of indifference in the quantum
world. The result is of course again Born's law. For me the Gleason
version and the envariance demonstration have the same value. They
both show that if one is going to introduce probability in the
quantum world, and therefore in the MWI, then  Eq.~2 is the most
natural choice. But still there is no ned for probability in the MWI
outside from experimental considerations foreign to the theory. At
that stage it is interesting to make a remark.
Everett~\cite{Everett} and later many authors such as Wallace and
Brown~\cite{brown,book,pattern}, or Zurek~\cite{Zurek} often claimed
that their bayesian definition of probability is no better nor worst
than in classical physics so that the situation is the same as for
the PWI. Everett for example, wrote~\cite{Everett}: \emph{The
situation here is fully analogous to that of classical statistical
mechanics, where one puts a measure on trajectories of systems in
the phase space by placing a measure on the phase space itself, and
then making assertions (such as ergodicity, quasi-ergodicity, etc.)
which hold for ``almost all'' trajectories.} By almost Everett means
here a measure-theoretic definition like the one proposed by
Lebesgue. However, having a measure is not enough to define a
probability. For example, we could use Noether symmetry theorem
which shows that Maxwell equations involve a conserved current and
interpret this result as a probability. Still, this is not necessary
,i.e., not required by the theory. We need a clear ontological
statement for introducing probability in a theory and this involves
dynamical elements like particle randomness. In the PWI for example,
the conservation law $\boldsymbol{\nabla}\cdot\mathbf{J} +\partial_t
\rho=0$ is clearly not enough to generate the probability
interpretation it only gives an indication. As a proof observe that
Madelung~\cite{madelung} found simultaneously with de Broglie the
guidance formula and the quantum potential. But he interpreted them
instead as continuous hydrodynamic fluid equations more in phase
with the old Schrodinger interpretation. Modern practitioners of the
MWI often think that by comparing their own approaches with the one
developed for instance by Gibbs using Liouville's theorem could give
a legitimacy to the various concepts of probability they propose.
However, Gibbs statistical mechanics is nothing without the physical
interpretation proposed by Boltzmann and Maxwell within the kinetic
theory (as Gibbs himself recognized  in the introduction to his
book~\cite{Gibbs}). Therefore, we should not try to extract too much
from
Gleason's theorem, decision theory or envariance.\\
\indent To introduce anyway probability we should love to the second
step of Everett's reasoning and consider a statistical experiment.
Suppose for instance a single quantum source emitting photons one by
one, all directed on a balanced beam splitter. Each photon is either
transmitted or reflected with equal probabilities and the detectors
(i.e., avalanche photodiodes) register singular events in only one
of the two path at once.  The statistics is of course a simple
Bernoulli process and the result, following the weak law of large
numbers, is naturally in agreement with Born's rule  that the number
of photons detected in each detectors are equal on the long run.
Now, this is of course a reading which make sense in Bohr's
instrumentalist interpretation as well as in the PWI. In the the PWI
photons follow trajectories determined by Eq.~3 (or some equivalent
\cite{drezet}) and by the initial conditions in the wave packets
(i.e. for example the particle's position at a given time). The
theory is very similar to classical mechanics and therefore the PWI
contains enough ingredients to introduce statistics in quantum
mechanics.\\
\indent  In the MWI this is not the case since the continuous wave
or field is all what we have. The wave packet impinging on the beam
splitter  behaves like a classical Maxwell wave: it is separated
into two equal parts but there is no  probability in this
construction. However, should we be surprised? The MWI is indeed
based on the pure unitary Schrodinger dynamics which only accepts
the regular solution of Eq.~1. There is no singularity, no fine
graining in this theory for generating randomness. It is therefore
surprising that Everett in his thesis \cite {Everett} followed by
B.~DeWitt and his student N~Graham \cite{stat} believed that
probability could appear in the long run in the MWI. It is of course
true  that in a Bernoulli process with $N$ repetition of the beam
splitter experiment there is $W(n_t)=N!/n_t!(N-n_t)!$ possible
branches in which $n_t$ photons are transmitted while $N-n_t$
photons are reflected.  Using Stirling formula $N!\approx N^N$ (this
is a crude approximation) in the long run limit $N\rightarrow+\infty
$ we get $W(\tilde{n}_t)\approx 2^N$ for $\tilde{n}_t/N=1/2$  which
means that the overwhelming majority of the Everett `branches' will
be following Born's law. Still, this has only  value in a
measure-theoretic sense.  We added branches to find a degree of
reality equal to $W(n_t)/2^N$, but nowhere  there is a reason why we
should do that. In the same way that typicality is not probability
there is here no possibility for extracting `a tend to from a
does'\cite{Deutsch}. Additionally, for practical experiments we
don't use an infinite number of registered events. The PWI can deal
with that only because  the theory is deterministic and because
randomness only results from the choice of the initial conditions
for the particle positions. There is no need for an infinite number
of events in the same way that in kinetic
theory a gas doesn't contain an infinite number of molecules (the Gibbs ensemble is just an idealization after all).\\
\indent I stress that my comments are somehow well
known~\cite{comment} and that we can still  find some proponents of
the MWI who are convinced that Everett's answer is the good one. In
the recent year D. Wallace and S. Saunders following the ideas of
D.~Deutsch~\cite{Everett,book} introduced bayesianism into the MWI
in order to save the theory (this is what they called decision
theory following L.~J.~Savage~\cite{savage}). Quoting D.~Wallace on
his discussion on patterns~\cite{pattern}: `\emph{we have at least
shown that it is rational for the observer to assign  some
weighting: in other words, we have shown that there is room for
probabilistic concepts (at least the decision-theoretic sort) to be
accommodated in the theory.} In other words, the observer doesn't
know where he will go after `branching' so that it seems legitimate
to call that a bayesian choice. Or is it not?  Bayesianism following
Ramsey or de Finetti is mainly a subjective interpretation of
probability based on inferences and degrees of belief as used by
poker players or insurance companies. Still, a poker player can only
use his subjective notion of probability in connection with
empirical evidences. Such empirical evidence means frequency of
occurrences (I.e., ideally using an infinite sequence following von
Mises cf. however our remarks made earlier concerning Gibbs's
ensembles \cite{foot}). Therefore we are sent back to our first
criticism concerning the Everett, DeWitt-Graham reasoning.\\
\indent  As we explained before the unitary Schrodinger evolution is
too regular and simple for implying objective probability and
randomness. And it is therefore difficult for me to understand how
some could even hope to obtain physical probability directly from
subjective bayesianism (my criticism concerns also the new trends
about Qbism even though Qbism is actually the Copenhagen counterpart
of the decision theoretic view on the MWI). Furthermore, I am
feeling that the whole game of decision theory is a bit like meeting
a pair of twins or clones (see the very suggestive `Zaxtarian'
scenario proposed by B.~Green~\cite{Green} based on the `sleeping
beauty' analogy by L. Vaidman~\cite{sleep}) and asking one of them
what was the probability to be the other. In that case, at least,
ontogenesis (the fine grained structures) make some difference but
with the quantum states the symmetry is too high and the
fine-graining (unlike in the PWI) absent. Why therefore should we
introduce probability? This is not convincing, and the problem is
not that the observer doesn't know where he will go but rather that
he will actually choose both ways since he will soon be sliced into
two quantum clones. The MWI is deterministic and there is no hidden
variable with unknown initial conditions which could tell you what
was the path you really followed in a quantum interferometer.
Therefore, how can we speak of some results being likely and others
being unlikely when both take place? In one sense the strong
symmetry required by Zurek's envariance, i.e., the quantum version
of Laplace's indifference principle, already ruins the chance of
success of the MWI in justifying the use of bayesian probability.
Over the years L. Vaidman often emphasized (as a kind of last chance
for the MWI?) that probability should be postulated in the
MWI~\cite{Vaidman}. If this is true, then we completely denature the
dream of Everett and therefore we will at the end obtain something
like a new version of the PWI and return to the original de
Broglie's proposal. In the following I will describe some
alternatives to the old MWI, which indeed, are going to explore this
analogy with the PWI.
\section{Saving the Many-Worlds?}
 \indent In the recent
years, several approaches have been proposed to modify or complete
the MWI. I already mentioned the various many-minds
interpretations~\cite{Albert}. Such approaches include in the theory
some new ingredients which look a bit like hidden variables in the
PWI (in some versions the `minds' are indeed flying over the
wave-function like the electron is surfing the guiding wave in the
PWI: this is a romantic view). I will not analyze here such
provocative ideas, but the point is, indeed, that one way of saving
the MWI is to go in the direction of the PWI. This is what is
interesting to me here since the proposal is actually going far
beyond the realm of the many minds interpretation(s). Indeed, the
analogy with the PWI is for example the path proposed by
A.~Valentini~\cite{book}, in his many Bohmian worlds theory. Such an
approach is also advocated by several authors like F.J.~Tipler and
K.~J. B\"{o}lstrom~\cite{others}. The main point is that the PWI
allows a description of the wave function of the  whole Universe by
adding a Gibbsian ensemble of Universes to the theory. Indeed, if we
consider and infinite number of copies of such a Universe all
characterized by the same wave function $\Psi$ but in which the
trajectories $X_\Psi(t)$, for the large vector associated with the
$N$ particles of the Universe in the configuration space, differ
only  by the choices concerning the initial conditions
$X_\Psi(t_0)$, then we have a theory where many Bohmian worlds exist
independently of each others. This reminds us of the old criticism
by D.~Deutsch about PWI as a MWI in a permanent state of chronical
denial. Here, the aim is clearly to satisfy both PWI and MWI
proponents by reversing the critics. By introducing many copies of
the Universe we could, maybe, understand how the quantum potential
can depend on the wave function itself. N.~Rosen~\cite{Rosen} who
was one of the first to study the PWI (even  before Bohm) gave up on
this theory because it was not acceptable for him that a quantum
potential $Q_\psi$ should depend somehow on the probability density
(remember that for a single non relativistic particle we have
$Q_\psi=-\frac{\hbar^2}{2m}\Delta|\psi(x)|/|\psi(x)|$). De Broglie
renounced his PWI in 1930 in part for similar reasons: if the wave
function is a `subjective' element (the word is from de Broglie),
then how could a dynamics depends on statistics? De Broglie came
back to PWI in 1952 since this problem was not anymore a serious one
for him.  Still, we can motivate the many Bohmian worlds view by
introducing the quantum potential as a kind of interaction between
the different worlds. Actually this is even the only justification
for the many Bohmian worlds view. Personally, I am not so much
convinced but the theory is fine to study anyway since we don't
know yet where it will lead us.\\
\indent To conclude this chapter, I would like to discuss a
different provocative proposals for modifying the MWI which I
imagined some years ago. Since I am not taking it too seriously I
will call that model the jumper interpretation (JI). In the JI the
idea is to find a stochastic approach based on the MWI.  I will
consider first a single electron  described by $\Psi(x,t)$. My
suggestion is that at any time there is a single particle in the
wave like in the PWI. The particle is located at $x$ at time $t$.
The density of probability that it will be found at $x'$ at any
other time is simply given by
\begin{equation}P_\psi(x',t')=|\Psi(x',t')|^2.\end{equation}
 That's more or
less all. The evolution is completely stochastic and if the wave is
going to be diffracted then the electron is jumping from one place
to a different one simply using Born's rule. As inPWI, there is an
obvious privileged basis here: the spatial coordinate representation
but we could choose a different one. Also, the dynamics can look
`crazy'. First, the electron at time $t'$ has a non vanishing
probability to appear at any place in the wave whatever the distance
separating the two points $x$ and $x'$ (the velocity $(x-x')/(t-t')$
can therefore diverge). Second, consider a particle interacting with
a beam-splitter. After crossing the apparatus we have now two
separated wave packets. Still, from my JI the particle can jump from
one beam to the other completely randomly and this even if there is
a huge distance between them and even if there is a thick wall as an
obstacle. The theory also requires a privileged Lorentz Frame (like
PWI).  Indeed since there is not limit for velocity the trajectories
are allowed to go backward in time in some reference frames. A
privileged slicing of space-time allows us to define unambiguously
the possible probability of reaching a point using Born's rule
$P_\psi(x',t')=|\Psi(x',t')|^2$ calculated in this preferred frame.
Actually, this model is a minimalist version of the PWI. Instead of
being deterministic it is stochastic. But still it shares with the
PWI many common features. In particular, in a state like
$|\textrm{here}(t)\rangle+ |\textrm{there}(t)\rangle$ the particle
is  in one place at a time if we choose to favor the $x$
representation. If we take as preferred basis the momentum
representation then the system will not have position at all but
rather a jumping momentum $k$ in the distribution $|\langle
k|\textrm{here}(t)\rangle+ \langle k|\textrm{there}(t)\rangle|^2$.
This JI could also be used to explain the EPR paradox by including
entanglement and by generalizing it to any possible Hilbert space
(for example to include spin or polarization variables). Consider
for example a perfectly entangled photon pair state like
\begin{equation}|H\rangle_1|H\rangle_2+|V\rangle_1|V\rangle_2\end{equation} (where $H,V$
denote horizontal and vertical polarization for photons 1 and 2). We
require a preferred basis for describing the stochastic dynamics,
let say, $|H\rangle_i$ and $|V\rangle_i$ (but the choice could be
different from pair to pair ). When the pair of photons is produced
by the source it jumps randomly from time to time into the states
$|H\rangle_1|H\rangle_2$ or $|V\rangle_1|V\rangle_2$. Now the EPR
pair is directed in the Bell experiment with polarizing beam
splitters  and wave plates (for selecting polarization bases). Due
to the presence at each time of both waves $|H\rangle_1|H\rangle_2$
and $|V\rangle_1|V\rangle_2$ the outcomes will follow the quantum
predictions.  The situation will be like for the PWI involving a non
locality, this time not due to the quantum potential  but to the
mere existence of the two independent
branches$|H\rangle_1|H\rangle_2$ and $|V\rangle_1|V\rangle_2$: one
full, one empty at each time.\\
\indent The JI can be  easily extended to measurement situations
like the one described by Eq.~5 with many electrons. Since we have
privileged some ket basis in this theory the quantum state has a non
ambiguous meaning in such a basis. This is again very similar to the
PWI. Here however the choice of the privileged basis is arbitrary: I
could have chosen a momentum space instead of a spatial coordinate
this is my free ontological assumption. Now, of course there is this
strange randomness induced by the jump of particles in the wave
function. What  is this jump going to imply for the brain states and
to the observers? Are they going also to jump or to see  these
electrons jumping?  If the electrons are jumping  in front of my
eyes then the theory is of course invalidated. But, this is again a
hidden variable model and actually the nice thing is that this jump
is really hidden (in the same way that Bohmian trajectories are
hidden in the PWI). Indeed, we see first that if we consider an
awareness state, like $\langle X_1,....X_N|\ddot{\smile}\rangle$
written in the preferred basis (the question of the basis choice is
again arbitrary but to fix it let's say that I am working mainly
with spatial coordinates + the spin degrees of freedom) then the
mind somehow supervenes on the $N$ particles position at a given
time (this is also true for the PWI).  Second, in a state like Eq.~6
\begin{equation}
\Psi(\mathbf{x}_1,X_1,....X_N,t)=\psi'_1(\mathbf{x}_1,t)\langle
X_1,....X_N|\ddot{\smile}_t\rangle+\psi''_1(\mathbf{x}_1,t)\langle
X_1,....X_N|\ddot{\frown}_t\rangle+...
\end{equation} there is  only one electron 1 at $\mathbf{x}_1$ and one
brain (whatever that means) located at $X_1,....X_N$. Like for the
PWI it doesn't matter if we use Eq.~6 with awareness states or Eq.~7
with cat states. In both the probability of presence is given by the
same number
$P_\Psi(\mathbf{x}_1,X_1,....X_N,t)=|\Psi(\mathbf{x}_1,X_1,....X_N,t)|^2$.
Now, the question about patterns plays an important role here and
discussing it now will also bring some new insights to the PWI. What
is important, indeed, is that in both the present JI, and in the
PWI, the whole system only occupies one position
$X=[\mathbf{x}_1,X_1,....X_N]$ at any time, only restricted by the
condition $P_\Psi(\mathbf{x}_1,X_1,....X_N,t)\neq 0$. Now, you
should remember that we are speaking about a quantum measurement and
that $\psi'_1(\mathbf{x}_1,t)$ and $\psi''_1(\mathbf{x}_1,t)$ are
not spatially overlapping (at least at some time $t$). It means that
we have either
\begin{eqnarray}
P_\Psi(\mathbf{x}_1,X_1,....X_N,t)=|\psi'_1(\mathbf{x}_1,t)|^2|\langle
X_1,....X_N|\ddot{\smile}_t\rangle|^2\end{eqnarray}or
\begin{eqnarray}
P_\Psi(\mathbf{x}_1,X_1,....X_N,t)=|\psi''_1(\mathbf{x}_1,t)|^2|\langle
X_1,....X_N|\ddot{\frown}_t\rangle|^2
\end{eqnarray}
depending on whether the electron position $\mathbf{x}_1$ is located
in the support of  $\psi'_1$ or $\psi''_1$. The functionalist theory
together with the existence of particles therefore imposes that we
can only be in one of the two awareness states! This trick is not
possible in the old MWI interpretation since there is no preferred
basis in Eq.~1. Additionally, it apriori gives a reason to impose
the spatial coordinates as preferred basis since using momentum with
its delocalized quantum states
$\tilde{\psi}'_1(\mathbf{k}),\tilde{\psi}''_1(\mathbf{k})$ could not
lead to such a
clean resolution of the observer identity crisis. This is, I think, a very good reason to prefer the PWI over the MWI.\\
\indent Now, with the present stochastic model the state given by
Eq.~5 represents an electron 1 and the detectors as jumping
together. The three  entangled electrons will jump from
$\psi_1'|g_2\rangle|e_3\rangle$ to $\psi_1''|e_2\rangle|g_3\rangle$
randomly between two arbitrary times. In the state given by Eq.~6
with an observer entangled with electron 1 the whole system should
also jump but, again,  only together. That means, that the observer
cannot feel this jump at all!  If he is jumping all his memory
sequence and the environment is jumping with him (this why I called
the theory crazy). Furthermore, if other observers are included in
the formalism (see Eq.~8) then the subjective agreement will also be
automatic in the appropriate basis and You will jump with Me always
in agreement with quantum mechanics. I emphasize that this JI is
certainly a schematic structure not to be taken too seriously; but I
think that it contains mainly all the ingredients for a good hidden
variable theory. In particular, the concept of probability used in
the model is clear enough to bypass all the contradictions contained
in the MWI.  As a further development we could also present a
many-jumpers interpretation similar in philosophy to the many
bohmian worlds approach discussed before. Again, this could bring
some philosophical advantages for explaining how the particles
`know' where to jump. The inter communication between the different
jumping worlds brings indeed such an information to the whole
ensemble of Universes and with this we will conclude our
`many-stories' theory.\\

The author thanks Nayla Farouki and Lev Vaidman for helpful
suggestions and comments during the preparation of the manuscript.


\begin{references}
\bibitem{foot}
D. Wallace often criticized the frequentist view using the well
known redundancy with the weak law of large numbers. Indeed, this
weak law is sometime used, improperly, as a proof of the long run
probability formula. However, since the weak law already needs
probability (with its frequentist definition)  to run there is
obviously a kind of circularity and this motivates Wallace's
bayesian reading of probability.  I however think that the naive
view that  the weak law is a proof  of something like the
frequentist interpretation is not exact. The frequentist view,
indeed, is not proven by the weak law of large number: it only shows
that this view is consistent with itself. Actually, the frequentist
view as often taught is too simple for exhausting the full physical
complexity of randomness. As I mentioned, Gibbs ensembles are pure
idealization and  what we need is actually a physical postulate
concerning the initial conditions of the Universe. If these
conditions impose that over $N\approx 10^{23}$ molecules $N_1$ will
be in state 1 and $N_2=N-N_1$ in state 2 then we have already a good
basis for probability. However, this is not enough in order to have
a good random Universe since in the long run a maverick result like
$00...11$ (with $N_1$ `$0$' arriving first and $N_2$ `$1$' after) is
not random but would however satisfy the naive frequentist view. In
my opinion, we must also add a physical postulate similar to
Boltzmann's \textrm{stosszahlansatz} telling that the repartition of
`$0$' and `$1$' in the finite sequence is somehow `random' as
determined by initial conditions. This postulate would for example
mean that the molecules with state 1 and 2 are homogeneously
distributed over the box or the Universe considered, and that the
observer should pick-up a fair sample when making a statistics.  I
agree that mathematicians will not like my postulates, but in my
opinion probability should be a physical science before being a
domain for mathematicians. In other words, probabilities are only
pure idealizations in a deterministic Universe and they only
approximately convey a language for what we call chance and
randomness. Unfortunately we don't have a better one.
\end{references}
\end{document}